%Paper: cond-mat/9502101
%From: Italo Marino <italo@bcs.uchicago.edu>
%Date: Thu, 23 Feb 95 20:04:57 CST

%%%%%%%%%%%%%%%%%%%%%%%%%%%%%%%%%%%%%%%%%%%%%%%%%%%%%%%%%%%%%%%
%%% 	TEX FILE  (02/22/95)
%%%%%%%%%%%%%%%%%%%%%%%%%%%%%%%%%%%%%%%%%%%%%%%%%%%%%%%%%%%%%%
%%%%%%%%%%%%%%%%%%%%%%%%%%%%%%%%%%%%%%%%%%%%%%%%%%%%%%%%%%%%%%
\magnification=\magstep1
\def\eq{$$}
\def\en{$$\medskip\noindent}
\def\hh{\hskip 1 true pt}
\hsize=6.0 true in
\vsize=8.5 true in
\voffset=0.5 true in
\baselineskip=24 true pt
\def\wj{\omega_J}
\def\Wac{\Omega_{AC}}
\def\Wj{\Omega_{J}}
\def\wac{\omega_{AC}}
\def\Iac{I_{AC}}
\def\iac{i_{AC}}

\def\Idc{I_{DC}}
\def\ad{\iac/\wac}
\def\ph{n \psi + m \wac \, \tau}
\def\cline{\centerline}
\nopagenumbers
\headline{\hfill}
\bigskip\bigskip\bigskip
\bigskip\bigskip\bigskip
\bigskip\bigskip\bigskip
\font\tenrm=cmr10 scaled\magstep3

\baselineskip=30 true pt
\cline{\bf {\tenrm Giant Shapiro Steps in Josephson Arrays:}}
%\cline{\bf {\tenrm  Two-Dimensional Frustrated Josephson Arrays:}}
\cline{\bf {\tenrm Analytical Results}}
\baselineskip=18 true pt
\font\tenrm=cmr10 scaled\magstep2
\bigskip\bigskip\bigskip
\cline{Italo F. Marino}
\smallskip
\cline{\it The James Franck Institute and The Department of Physics,}
\cline{\it The University of Chicago, Chicago, Illinois 60637}
\smallskip
\bigskip\bigskip
\baselineskip=19 true pt
\cline{\bf {\tenrm Abstract}}
\medskip
We study two-dimensional  Josephson arrays driven by
a combined DC plus AC current, and with an applied transverse
magnetic flux of $f$ flux quanta per plaquette.
We present {\it ansatz} solutions
for sufficiently large  frequencies, which  are a generalization
of the travelling wave solutions
found  by Marino and Halsey$^{8} \,$ for the case of
 DC current driving.
For $f=1/2$ and $f=1/3$, we compute the  widths of the first
few Shapiro steps for both integer and fractional winding numbers.
 These
 expressions consist of products of Bessel functions of
$\, (\ad)$, where $\iac$ and $\wac$ are the amplitude and the
frequency of the driving AC current, respectively,
times a frequency-dependent factor for fractional steps.
In the limit of large  frequencies,  we find that the fractional steps
are suppressed, whereas the maximum integer step widths
   saturate to a frequency-independent value.
 We show that  the suppression of the fractional steps
is due to decrease of the vertical
(i.e. perpendicular to the direction of flow of the injected
current) supercurrent relative to the
normal current, whereas the persistence of the integer steps is
due to the existence of zero-frequency (though spatially varying)
terms in the expansion for the gauge-invariant phase differences,
for which the normal current vanishes.
 These results are in reasonable  agreement with the numerical computations
carried out by other groups.$^{3,5}$

\vfill
\line{PACS numbers: 85.25.Cp, 74.50.+r, 05.45.+b  \hfill  02/22/95}
\eject
\baselineskip=24 true pt
\bigskip
\noindent
{\bf I. INTRODUCTION}
\medskip
\headline{\hfil \folio \hfil}

When a resistively shunted Josephson junction is driven by a combined
DC and AC current $I(t)=\Idc+\Iac \sin (\Wac \hh t)$, the current-voltage
characteristic exhibits plateaus in which the time-averaged voltage
is  equal to an integer multiple of $\, \hbar \Wac/2 e \,$ for a
finite interval of DC current. These plateaus are called {\it Shapiro
steps}.$^1 \,$  An analogous effect is observed when the current is applied
to
an $N \times N$ square array of Josephson junctions with  tranverse
magnetic flux per plaquette $\Phi=f \Phi_0$, where $\Phi_0=\hbar c/2 e \,$
is the quantum flux, and $f=p/q$ is  the frustration,  $p$ and $q$
being  relatively prime integers. The total voltage across the
array is locked at values given by

\eq
V_{N}={n \over q} {N \hbar \Wac \over 2 e}.
\eqno(1)
\en
In this case, the steps are called {\it fractional giant
Shapiro steps}.$^2 \,$  The accepted explanation of this
effect is that the $\, q \times q \,$ periodic vortex super-lattice
moves coherently in response to the external AC field.$^{2,3}$\par

If the parallel shunt resistance is $R$, and the critical
current per junction is $i_0$,
 we can then define a dimensionless time $\tau = (2 e  i_{0} R/\hbar)\, t$.
Measured in these time units, the external AC frequency is
$ \wac \equiv \hbar \Wac/2 e  i_{0} R$.
The  Josephson frequency is defined by $ \Wj \equiv 2 e V_{N}/
N \hbar$, with its normalized version  given by
$ \wj \equiv V_{N}/i_{0} R N$.
In terms of them, the above relation can simply be expressed as
$ \wj=\nu \hh \wac$,
where $ \nu \equiv n/q$ is  the winding number.\par

These steps display a variety of characteristics.
It has been observed that there exists a qualitative difference between
the cases where $ \nu$ is an integer ({\it integer steps}), and when
it is a fraction
({\it fractional steps}).$^{2-5} \,$ This difference becomes manifest when
going from low to high frequencies. At low frequencies, both fractional
and integer steps
behave qualitatively as in the single-junction case. In the high-frequency
limit, on the other hand,   fractional steps are suppressed, whereas
the maximum integer step widths  saturate to a
frequency-independent value.$^{4-6}$\par

This phenomenon has been widely studied by several authors.$^{2-7} \,$
Analytical treatments have been provided by Halsey,$^7$ Lee and
Halsey,$^6$ and by Rzchowski et al.$^4$
 Nevertheless, a theoretical derivation based on first principles
is still lacking. The missing link in the understanding of
this problem has been the knowledge of
 the solutions  for  DC plus AC current driving.
Progress in this direction has already been made.
The existence of a family of travelling wave solutions
for  DC current-driven arrays has been
reported by Marino and Halsey$^{8}$ in the limit of high
Josephson frequencies.
In this paper we present a generalization of these travelling
wave states to the case where the array is driven by an
additional AC current. For $f=1/2$, these are also solutions
to the model equations used by Rzchowski et al.$^4$ In addition to
the modes corresponding to the Josephson frequency $ \wj $ and its
harmonics,
these solutions contain terms oscillating with
 frequencies  given by linear combinations
of $\, \wj \,$ and $\, \wac \,$.
If one then computes the  current flowing across
the entire array, one finds that only the terms  with
frequencies
given by $\, (k_1 q \, \wj + m \, \wac) \,$ survive, where $k_1$ and $m$
are integers. This  is due to the fact that these terms
are exactly on phase everywhere on the array, whereas the other terms
have phases such that their sums vanish.
Shapiro steps  result when this linear combination
of frequencies is zero.
We then derive  expressions for the step widths as a function
of $\, \iac \,$ and $\, \wac \,$ for $f=1/2$ and $f=1/3$, by computing
the ensuing DC supercurrent corresponding to this mode across the array
  for the first few steps.
These expressions consist  of products of $q$ Bessel functions
of $\, (\ad)$,  and other factors that depend  solely on the frequency
and an arbitrary constant phase $\psi_0$.\par

 The main idea ensuing from  this analysis is that
  the  difference between the behaviors of
the fractional and integer steps at low and high frequencies is
determined by  the relative size of the vertical $\,$(perpendicular
to the direction of flow of the injected current) supercurrent
relative to the normal current for the different frequency
modes contributing to a given step width.
  At low frequencies the vertical supercurrent is dominant for all
modes, and hence,
both integer and fractional steps  behave  in the same manner.
 When the frequency is increased, all the modes with non-vanishing
normal currents decrease.
The suppression of the fractional steps at high frequencies
appears as a consequence of their dependence on these modes,
 whereas the persistence
of the integer steps is due to the existence of zero-frequency
 modes (and therefore, with vanishing normal currents) whose
amplitudes are determined by the vertical supercurrent. Consequently,
integer steps behave  in the same fashion in both the
low- and high-frequency regimes. This is the same mechanism that yields
the Shapiro steps for a single Josephson junction.\par

Our solutions for the gauge-invariant phase differences contain the
arbritrary phase $\psi_0$.
On a given step, this phase varies over an interval
that we hypothetise to be frequency-dependent. At high frequencies we
shall assume  that this interval attains  a constant size, which
in principle can only be determined by fitting the
numerical results to the theoretical predictions.
Our ignorance about the details of the nature
 of the solutions at low
frequencies does
not allow us to provide expressions for the step
widths in this regime.
Our study, thus, does
not address the problem of trying to find the true  variation of this
interval with the frequency.\par
In section {\bf II} we review  the
solutions for the DC case and present
their generalization to the case of DC plus  AC  current driving.
In section {\bf III}
we use these
solutions as starting point to compute the widths of the first few
Shapiro  steps for $f=1/2$ and $f=1/3$. Finally, in section {\bf IV} we
provide  results from numerical computations.\par
\vfill
\eject
\bigskip
\noindent
{\bf II. SOLUTIONS FOR DC PLUS AC CURRENT DRIVING}
\medskip
We consider a square array of $N \times N$ overdamped resistively shunted
Josephson junctions in a uniform transverse magnetic field with $f$
 flux quanta
piercing each plaquette, parallel shunt resistance $R$, and critical
current per junction $i_0$. We define the gauge-invariant phase
differences by $\theta_{ij} \equiv \theta_i -\theta_j - A_{ij}$,
where $\theta_i$ is
the superconducting phase on the $i$'th site on the array, $\, A_{ij}$ is
the line integral of the magnetic vector potential $\, A_{ij}=
{2 \pi \over
\Phi_0} \int_{i}^{j} \vec{A} \cdot \vec{dx}$, such that
$\sum_{P} A_{ij} = 2 \pi f$, where the sum is around a plaquette,
and $j$ denotes a site that is  nearest
neighbor with $i$.
Then the current flowing from the $i$'th site to
the $j$'th site is

\eq
\tilde{I}_{ij} = {d \over d\tau} (\theta_{ij})+
\sin(\theta_{ij}),
\eqno(2)
\en
where $\tilde{I}_{ij} \equiv I_{ij}/i_{0}$. The first term
is the normal current, while the second one represents the
supercurrent.
The equations of motion simply express
the fact that the total current arriving at each site on the array
should equal the current externally injected there

\eq
\sum_j \tilde{I}_{ij}=i_{i;ext},
\eqno(3)
\en
where  the external current $\, i_{i;ext} \,$   vanishes everywhere
inside the array, except for at the boundaries.
 Henceforth, we shall take the convention that
when computing the gauge-invariant phase differences, on
horizontal bonds $j$ is to be taken to the right of $i$,
and above it on vertical bonds.\par

For the case in which the system is driven by a uniform DC current
injected parallel to one of the axes of the array (which we take
to be the horizontal axis with the current flowing from right to left),
and with periodic boundary conditions along the vertical direction,
Marino and Halsey$^{8}$ reported the
existence of a family of
travelling wave solutions. These solutions are characterized by a
parameter $\delta$ that measures the phase shift of the
phase oscillations along the horizontal direction. Along
the vertical direction the phase
shift is simply equal to $2 \pi f$, which is consistent with
the condition of transverse periodic  boundary conditions with period $q$.
These solutions possess a combined spatio-temporal translational
symmetry, in the sense that a translation of the solution by
one lattice spacing along the horizontal direction is equivalent to
the translation of the solution by a time $\tau = \delta /\wj$,
and a translation of the solution by one lattice spacing along
the vertical direction is equivalent to a translation of the
solution by a time $\tau = 2 \pi f/\wj$.\par
 Let us define

\eq
\psi\equiv \wj \hh t + 2 \pi f \, n_{Y} + \delta n_{X} +\psi_{0},
\eqno(4)
\en
where
$n_{X}$ and $n_{Y}$ are integers, and $\psi_0$ is a constant phase.
Then, the  gauge-invariant
 phase differences
on horizontal and vertical bonds for these solutions are  given by
\eq
\theta_{H}(\psi)= \psi + f_{H}(\psi),
\eqno(5)
\en
\eq
\theta_{V}(\psi)=f_{V}(\psi),
\eqno(6)
\en
where $f_{H}$ and $f_{V}$ are periodic functions with period $2 \pi$
and zero time
average. The authors of reference $8$ worked out the analytical form
of these functions in the limit of high voltages, by retaining only
the first harmonic in their Fourier expansion.\par

The generalization of these solutions to the case of
combined DC and AC current driving
is straightforward. The main modification is that now in addition to
the mode with Josephson
frequency $\, \wj$, a mode with frequency $\, \wac \,$ should also
be present, due to the external AC current.
The presence of this second time scale ruins the spatio-temporal
translational symmetry in the cases in which $\, \wj \,$ and
$\, \wac \,$
are incommensurate. This entails no problem, as we shall see.
The beating of these two  modes
due the horizontal supercurrent requires the
additional presence of terms
with frequencies given by linear combinations of $\wj$ and $\wac$
in the expansion for the gauge-invariant phase differences.
These solutions
must further satisfy several conditions.  Firstly, they should  reduce
to their DC counterparts when $\, \iac$ and $\, \wac \,$ are set
equal to zero; secondly, in order to (indirectly) enforce
the boundary conditions,   the DC and AC currents
(at $\wac$ and its harmonics) $\,$ flowing on each horizontal bond
should be
independent of position, and equal to the values given by
the currents injected at the boundaries, whereas on vertical
bonds they should vanish, allowing
for the possible existence of zero-frequency modes (that occur
only on integer Shapiro steps), that
should not be regarded  strictly as DC currents;
and finally, the linear term should remain unchanged, because
it is still true that the slope should yield the average voltage per
junction.
Our {\it ansatz}~ for the gauge-invariant
phase differences on horizontal bonds then takes the form

\eq
\theta_{H} (n_X,n_Y,\tau)= \psi + \sum_{n,m} \, \alpha^{H}_{n,m}
\cos (n \psi + m \wac \, \tau+ \xi^{H}_{n,m}),
\eqno(7)
\en
\smallskip
\noindent
where $n \,(\ge 0)$ and $m$ are integers, while on vertical bonds we have

\eq
\theta_{V} (n_X,n_Y,\tau)=\sum_{n \neq 0,m} \alpha^{V}_{n,m} \cos (n \psi + m
\wac \, \tau + \xi^{V}_{n,m}),
\eqno(8)
\en
\smallskip
\noindent
where $\psi$ is given by Eq. (4), and  $\xi_{n,m}^{H,V}$ are constant
phases. The  phase differences
are taken according to our convention.
The different terms in this expansion are labeled by the two
integers $n$ and $m$. We shall refer to this component of the
phases and currents as the $(n,m)$ mode.
Notice that the modes with $n=0 \,$ have no
spatial dependence, in agreement with our assumption.
This form of solution is good enough
to describe the system even at low frequencies.\par

The gauge-invariant phase differences are not independent.
The sum of their oscillating parts around a plaquette has to
vanish. We impose this condition
to each frequency mode and obtain

\eq
 \alpha^{V}_{n,m}= \beta \, \alpha^{H}_{n,m},
\eqno(9)
\en
\eq
\xi^{V}_{n,m}=\xi^{H}_{n,m} + n \, (\pi f - \delta/2),
\eqno(10)
\en
where $\beta = \sin n \pi f/\sin (n \delta/2)$, if $n \neq \dot{q}$
($n = \dot{q} \,$ is a shorthand  for $n = q \,$ mod $0$),
and $0$ otherwise. Thus, the modes with $n=\dot{q}$ are absent
on vertical bonds. This result
is the same one  that was obtained
in the  DC case. The reader is refered to reference
$8$ for the details of the derivation.\par

On the  Shapiro steps we will assume that
$\delta = 2 \pi f$, due to our
requirement of AC translational invariance (see the discussion
preceeding Eq.~(7)). This implies
$\beta=1$ for $n \neq \dot{q}$, in which case  $\alpha$
and $\xi$ are the same on both horizontal and vertical
bonds, which considerably simplifies matters.
Consequently, we shall hereafter drop the superscripts
$H$ and $V$ in our expressions.\par

We shall perform
a mode  expansion of the horizontal supercurrent $\sin \theta_H$
 in the  following manner:

\eq\eqalign{
\sin \bigl(\psi + \sum_{n,m}
\alpha_{n,m} & \cos \bigl(n \psi + m \wac  \tau + \xi_{n,m})
\bigr) \cr
= & \sum_{n,m} S_{n,m} \cos \bigl( n \psi + m \wac  \tau +
\Xi_{n,m} \bigr). \cr}
\eqno(11)
\en

The different components of the supercurrent can be computed
in a straightforward manner by performing a Fourier-Bessel
expansion of the left-hand side of Eq.~(11). A generic
term in this expansion is of the form

\eq\eqalign{
S_{n,m} \cos (n \, \psi +   m \, \wac & \, \tau +  \Xi_{n,m}) \, =
 \sum_{\lbrace (n_j,m_j) \rbrace } \Im \,  \biggl \lbrace \biggl( \prod
J_{k_j}(\alpha_{n_j,m_j}) \,
i^{k_j} \biggr) \cr & \exp \biggl(i \psi + i \sum_j \ k_j \,
(n_j \psi + m_j \wac \tau + \xi_{n_j,m_j})  \biggr)
\biggr \rbrace. \cr}
\eqno(12)
\en

The sum is over the set of sets of pairs $\, \bigl \lbrace
\lbrace (n_j,m_j) \rbrace \bigr \rbrace \,$  such
that there exists a set of coefficients $\lbrace k_j \rbrace$ for
which the following relationship (understood as a vector identity) holds

\eq
(n,m) = (1,0)_L + \sum_j \ k_j (n_j,m_j),
\eqno(13)
\en
where $\, (1,0)_L$ denotes the contribution due to the linear
term, which is absent on vertical bonds.
These expressions are in general quite complicated. We shall
assume that they can approximately be computed
starting from
the lowest-order (in $n$ and $m$) modes, since the magnitude
of the different Bessel functions decay quickly with
increasing order. This approximation should be good enough
at high frequencies, but we do not expect it
to remain accurate for lower frequencies. The feedback of the
higher-order modes on the lower ones should become more important
as one approaches the critical current.
On vertical bonds the $\, (n,m) \,$ component of
the  supercurrent is (neglecting higher-order corrections)
equal to $\, 2 J_1(\alpha_{n,m})$. It can be shown for
simple cases that
other contributions vanish.\par

We now turn to the equations of current conservation $\,$ (Eq.~(3)).
Once again, we have to distinguish between
the cases  $\, n \neq \dot{q} \,$ and $\, n = \dot{q} \,$. In the former
case, the equation for current conservation for our {\it ansatz}~
solution is

\eq\eqalignno{
-2 \omega_{n,m}  \alpha_{n,m}  \sin & (\ph + \xi_{n,m})
+ 2 J_{1}(\alpha_{n,m}) \cos (\ph + \xi_{n,m}) \cr & + S_{n,m}
\cos (\ph + \Xi_{n,m}) = 0,&(14)\cr}
\en
where
$\, \omega_{n,m}=n \wj + m\wac$. There is an overall factor of
$2 \sin \pi f$ that goes away. The first
term in the above equation represents the combined effect of both the
horizontal and vertical normal currents, each of them contributing
the same amount; the second one is due to
the vertical supercurrent, and the last one comes from
the horizontal supercurrent, which has to be computed for each
mode.
For $\, n=\dot{q} \,$ the
current is trivially conserved at each site. Furthermore, the
current corresponding to the   $\, (0,1) \,$ mode should equal the
external AC current:

\eq
-\wac  \, \alpha_{0,1} \, \sin (\wac \, \tau + \xi_{0,1}) +
S_{0,1} \cos (\wac \, \tau +
\Xi_{0,1})
=-\iac \sin (\wac \, \tau).
\eqno(15)
\en
For convenience, and without loss of generality,
we have introduced a minus sign at the right hand side of this equation.
If we neglect the vertical supercurrent
and replace the factor of $2$ multiplying the first term
in  Eq.~(14) by $1$, then these two last equations
describe an overdamped  single junction.\par
In general, we expect   $\, \alpha_{0,1} \sim
\iac/\wac$,  and  the asymptotic behavior of the different
amplitudes at  high and low frequencies to be of the form

\eq
\alpha_{n,m} \sim h(\wj)
 \prod_{\sum n_i \, k_i = m} J_{n_i}^{k_i}(\alpha_{0,1}).
\eqno(16)
\en

\noindent
for $\, n \neq \dot{q}$, where  $\, h(x) \sim  c \, (c = constant \le 1)$,
as $x \to 0$,
and $\, h(x) \sim x^{-n}$, as $x \to \infty$.
This will be made
clear below. Consequently,
 it is  safe to assume (save for $\, \alpha_{0,1}$) that
$\, J_{1}(\alpha_{n,m})
\approx \alpha_{n,m}/2$. Eq.~(14) can then be
 solved in terms
of the expression for the horizontal supercurrent:

\eq
\alpha_{n,m}=  {S_{n,m} \over \sqrt{4 \omega_{n,m}^2 +1}},
\eqno(17)
\en
\eq
\xi_{nm}=\pi + \Xi_{n,m} - \arctan (2 \omega_{n,m}).
\eqno(18)
\en

For $\, \omega_{n,m} \gg 1$ the effect of the vertical supercurrent
can be neglected and $\, \alpha_{n,m} \sim S_{n,m}/(2 \omega_{n,m})$,
whereas for $\, \omega_{n,m} \ll 1$, the vertical supercurrent
dominates and $\, \alpha_{n,m} \sim S_{n,m}$. Eq. ~(16) can then
be proven in the following manner.
 Since $\, (0,m) = (1,0)_L + m \ (0,1)$, then, to leading order
 $\, S_{0,m} = J_{m}(\ad)$.
 From Eq.~(17) it follows that $\, \alpha_{1,m}$ has the  asymptotic
behavior given by Eq.~(16). The proof for $\, \alpha_{n,m}$
 can be made by induction.\par
In the present calculation we will neglect the supercurrent
 in Eq.~(15). This is a good approximation for $\wac~\gg~1$.
 Thence,
 $\, \alpha_{0,1}=\ad \,$ and $\, \xi_{0,1}=0$. Using this,
 $\, (1,0) = (1,0)_L + 0 \ (0,1)$, and $J_{0} (\alpha_{n,m})
\approx 1 \,$ for all the other modes, we find

\eq
\alpha_{1,0}= {J_{0}(\ad) \over \sqrt{4 \wj^2 +1}},
\eqno(19)
\en
\eq
\xi_{1,0}={\pi \over 2} - \arctan (2 \wj).
\eqno(20)
\en
This solution also holds in the DC case $(\, \iac=0)$, and represents
a generalization of the solutions presented in reference $8$.
Unlike them, this solution remains regular as $\, \wj \to 0$,
owing to the vertical supercurrent.
Similarly, $\, (1, \pm 1) = (1,0)_L \pm (0,1)$.
\vfill
\eject
\noindent
Then,

\eq
\alpha_{1,\pm1} = {J_{1}(\ad) \over \sqrt{4 (\wj \pm \wac)^2 +1}},
\eqno(21)
\en
\eq
\xi_{1,\pm1} = \pi - \arctan (2(\wj \pm \wac)).
\eqno(22)
\en

\noindent
Other modes can be computed in a similar fashion.
\par
\bigskip
\noindent
%%%%%%%%%%%%%%%%%%%%%%%%%%%%%%%
%%%	THE SHAPIRO STEPS   %%%
%%%%%%%%%%%%%%%%%%%%%%%%%%%%%%%
{\bf III. THE SHAPIRO STEPS}
\medskip
It is clear that both the normal current and the
supercurrent have the same harmonic dependence as the
gauge-invariant phase differences. In particular,
this implies that when computing the total voltage and
supercurrent across the array, only the  modes with $\, n=k_1 q$
survive. It is immediate to check that all the other terms
 cancel out. This is usually interpreted in terms of
the vortex configuration by saying that a vortex moves $q$ times during
a period of $\, 2 \pi/\wj$.  This is key
in order to understand the phenomenon of the Shapiro steps. \par

Looking back at Eq.~(11), in presence of an external AC
current, an additional DC supercurrent
will appear across the array for frequencies satisfying
$\, k_1 q \wj + m \wac=0$, with  $k_1$ and $m$ relative primes. The
Shapiro steps  ensue.
We  see that the proposed rigid motion of the $\, q \times q \,$
vortex
 super lattice in response to the external AC field has a very
natural explanation within our theory.
For $k_1 > 1$ we have subharmonic steps. We will not consider
this possibility here, because these steps are in general too
small to be observed.\par

The first step in our calculation is to identify
the modes  that yield the largest contribution to the DC supercurrent
corresponding to the different Shapiro steps.
For integer steps $\, (\nu = n)$ the choice is unambiguos:
it is the
set of zero-frequency  modes $\, (k',-k' n)$, where $\, 1 \le \, k' < q$.
These modes have the virtue that their associated normal currents
vanish, and thus, they are determined by the vertical supercurrent.
The Shapiro step widths in this case
depend  only on $\, (\ad)$.
 The case of fractional
steps $\, ( \nu = n/q )$
can be analyzed
in a similar manner. The most important contributions are due to
the modes $\, (n',n'')$, with $\, n' \le n$ and $\, n'' < q$.
The amplitudes of these modes decay with the frequency because
they are mainly determined by the normal current.\par

The distinction between the low- and high-frequency behaviors just
amounts to saying that at low frequencies both fractional
and integer steps are in a supercurrent- dominated regime,
whereas at high frequencies only the integer steps are,
thanks to the zero-frequency modes.
Saying that the steps display single-junction
behavior is just another way of rephrasing this fact.\par

All of our  expressions  depend on the arbitrary
phase $\psi_0$.
In particular, the DC supercurrent on a given step depends on
this phase, and thus the width of the step will depend on
 the range of variation of it. The interval of variation of this
phase should depend on the dynamical stability of these solutions
and also on the nature of the solution for the $(0,1)$ mode
 at low frequencies
(recall that we neglected the supercurrent term in Eq.~(15)).
We conjecture that the size of this interval varies with the frequency,
growing from zero to an interval of constant size at high frequencies,
and that this is  the mechanism underlying the growth and saturation
of the steps. This assumption seems to be good enough to reproduce
all the observed
qualitative features of the steps. The quantitative correctedness
of our expressions will depend on the goodness of our guess
for the variation of this phase. According to this, our results
for a given frequency should differ at most by a constant
factor (for all values of the $\iac$) from the results
obtained from simulations or experiments.
We shall not attempt to resolve this issue here, rather,
we shall use a generalization of an {\it ansatz}~ used by Halsey,$^7$
that seems to work well to reproduce the observed numerical results
in a reasonable manner.
Our results have a factor of the form $\cos \,(q \psi_0 + \phi_{q,\nu})$.
We shall
assume that at high frequencies
$\, (q \psi_0 + \phi_{q,\nu})$ is centered at $\pi/2 \,$ for $q \,$ even,
and at $0$ for $q$ odd. Furthermore, for both $\, q$ even and odd, we
shall assume that this phase varies over the interval
$\, [-\pi/2,\pi/2 \,]$
for fractional steps, and over $\, [-\pi/2q,\pi/2q \,]$ for
integer steps.  We shall thus  restrict ourselves to making
predictions for the values of the step widths only for large enough
frequencies.\par

We now turn to specific examples. We shall only consider the cases
$f=1/2$ and $f=1/3$, because the number of modes that have to be included
in a given calculation increases quickly with  $q$.

\bigskip
\noindent
%%%%%%%%%%%%%%%%%%%%%%%%%%%%%%%
%%%	INTEGER STEPS    %%%%%%
%%%%%%%%%%%%%%%%%%%%%%%%%%%%%%%
{\bf a. Integer Steps: $\nu = n$}
\bigskip
\noindent
{\bf (a1) $f=1/2$}
\medskip
 A DC supercurrent occurs at the
mode $\, (2,-2n) = (1,0)_L + (1,-n) - n \, (0,1)$.
In this case $\, \alpha_{1, -n} = J_{n}(\ad)$, and $\, \xi_{1, -n} = (n +
1)\pi/ 2$.
We thus obtain the   expression:
\eq\eqalignno{
i_{1/2,n} =  & \,  J_{1}(\alpha_{1,-n}) \, J_{n}(\alpha_{0,1})
\sin (2 \psi_{0} + (2 n+1) {\pi/ 2}) \cr
= & \, {1 \over 2} \ (J_{n}(\ad))^{2}
\sin (2 \psi_{0} + (2 n+1) {\pi/ 2}),&(23)\cr}
\en
\noindent
where $\, i_{1/2,n}$ denotes the ensuing DC supercurrent. This
is not yet the expression for the step width.
The final answer depends on the range of variation of
$\psi_{0}$. Using our {\it ansatz}~ for this variation
at high frequencies we find

\eq
 \Delta i_{1/2,n}  = {\sqrt{2} \over 2} \, (J_{n} (\ad))^{2}.
\eqno(24)
\en
\vfill
\eject
\bigskip
\noindent
{\bf (a2) $f=1/3$}
\medskip
Here we have two contributions, namely,
$\, (3,-3n) = (1,0)_L + (2,-2n) - n \, (0,1)$, and $\, (3,-3n) = (1,0)_L
+ 2 \, (1,-n) - n \, (0,1)$.
Now, $\, \alpha_{2,-2n} = (J_n (\ad))^2
/2$, and $\, \xi_{2,-2n} = (n + 3/2) \, \pi$.
The total DC supercurrent corresponding to this step is

\eq\eqalignno{
i_{1/3,n}   = & \,  J_{n}(\ad) \, J_{2}(\alpha_{1,-n})
\sin (3 \psi_0 + 3 n \pi/2)
\cr  & {\hskip 8 pt} + J_1 (\alpha_{2,-2n}) J_{n}(\ad)
\sin (3\psi_0 + 3 n \pi/2) \cr  =  \, & {3 \over 8}
\, (J_{n} (\ad))^3 \sin (3 \psi_0 +3 n \pi /2).&(25)\cr}
\en
At high frequencies this becomes
\eq
\Delta i_{1/3,n} = {3 \over 16} (J_{n}(\ad))^{3}
\eqno(26)
\en
In conclusion, we find that the integer step widths are independent of the
frequency.
It  should be clear that in general
$\, i_{p/q,n} \sim (J_{n}(\ad))^q$. For $q~=~1$, this reduces to
the single-junction result, or equivalently, to the result for
the unfrustrated case $f=0$.\par
\bigskip
\noindent
%%%%%%%%%%%%%%%%%%%%%%%%%%%%%%%%%%%%%
%%%	FRACTIONAL STEPS      %%%%%%%
%%%%%%%%%%%%%%%%%%%%%%%%%%%%%%%%%%%%%
{\bf b. Fractional Steps: $\nu = n/q$}
\medskip
We need to compute the $(q, -n)$ component of the horizontal
supercurrent. As mentioned above, the most important contributions
come from the modes $\, (n',n'')$ such that $\, n' < q$ and
$\, n'' \le  n$.
The number of modes that are relevant for the determination of the steps
grows quickly with $n$, though, so we will only work out explicitly a
few simple cases.
\vfill
\eject
\bigskip
\noindent
{\bf (b1.1) $f=1/2, \, \nu=1/2$}
\medskip
We can either write $\, (2,-1) = (1,0)_L + (1,-1) +
0 \, (0,1)$,
or $\, (2,-1) = (1,0)_L + (1,0) - (0,1)$.
Omitting some details we get

\eq\eqalign{
i_{1/2,1/2} =  & -J_0(\alpha_{0,1}) J_{1}(\alpha_{1,-1})
\cos (2 \psi_0 +\arctan \wac) \cr &
- J_{1}(\alpha_{1,0}) J_{1}(\alpha_{0,1})
\cos (2 \psi_{0} - \arctan  \wac) \cr}
\eqno(27)
\en

\noindent
Using the expressions for $\, \alpha_{1,0}$, $\, \alpha_{0,1}$,
 and $\, \alpha_{1,-1}$, this turns into

\eq
i_{1/2,1/2} =  - {J_0 (\ad) J_1 (\ad) \over  (\wac^2 +1)} \,
  \cos \bigl(2 \psi_0 \bigr).
\eqno(28)
\en
and at high frequencies
\eq
\Delta i_{1/2,1/2} = {2 \, J_0(\ad) \, J_{1} (\ad) \over (\wac^2 +1)}
\eqno(29)
\en

We find that the step width decays like $1/\wac^2$ at high frequencies,
in disagreement with what has been assumed by other authors.$^{4,5}$
At low frequencies, this result reduces to a frequency-independent
expression, which is characteristic of single-junction
behavior. It is also
 possible to compute higher-order corrections (in Bessel
functions of $\, (\ad)$) by considering the combinations of modes
$\, (2,-1) = (1,0)_L + (1,1) - 2 \, (0,1) = (1,0)_L + (1,-2) + (0,1)$.
It is not hard to see that these corrections go like $\, J_1 (\ad) \,
J_2 (\ad) /(9 \wac^2 +1)$. This is negligible compared to the expression
given in Eq.~(29) for most cases of interest.
\bigskip
\noindent
{\bf (b1.2) $f=1/2, \, \nu=3/2$}
\medskip
In this case we have $\, (2,-3) = (1,0)_L + (1,-1) - 2 \, (0,1)$,
and $\, (2,-3)  = (1,0)_L + (1,-2) + (1,-1) + 0 \, (0,1)$. We will neglect
other contributions. The calculation is identical to the previous
one and yields

\eq\eqalign{
i_{1/2,3/2} = \,  & {J_1(\ad) \, J_2(\ad) \over  (\wac^2 +1)}  \,
 \cos (2 \psi_0 ). \cr }
\eqno(30)
\en

In the limit of high frequencies
\eq
\Delta i_{1/2,3/2} = {2 \, J_{1} (\ad) J_{2}(\ad) \over (\wac^2 +1)}
\eqno(31)
\en
\noindent
The next-order correction varies like $J_0 (\ad) \, J_3 (\ad)$,
which can be neglected.
We see that  this step width has the same dependence
on $\, \wac \,$ as for $\nu = 1/2$.
\bigskip
\noindent
{\bf (b2.1) $f=1/3,\,  \nu = 1/3$}
\medskip
This calculation involves higher-order modes than the ones hitherto
used . The number of terms, therefore,
considerably increases. In fact, we now have the possible combinations:
$(3,-1) = (1,0)_L + (1,-1) + (1,0) + 0 \, (0,1)$,
$\, (3,-1)~=~(1,0)_L + 2 \ (1,0) - (0,1)$,   $\, (3,-1)~=~(1,0)_L
 + (2,0) - (0,1)$, and $\, (3,-1) = (1,0)_L + (2,-1) + 0 \, (0,1)$.
There are two contributions to the $\, (2,-1) \,$ mode,
 which will be considered
separately. These are $\, (2,-1) = (1,0)_L + (1,-1) + 0 \, (0,1)
= (1,0)_L + (1,0) - (0,1)$. Putting
everything together we obtain:

\eq\eqalign{
i_{1/3,1/3} = & {J_{0}^{2}(\ad) J_1(\ad) \over 4 \sqrt{
4 \wac^2 /9 +1 }}
\biggl({\cos (3 \psi_0 -  2 \arctan (2 \wac/3)) \over
2 \sqrt{4 \wac^2 /9 +1 }} \cr &
+ {\cos (3 \psi_0 + \arctan (2 \wac/3) + \arctan (4 \wac/3))
\over \sqrt{16 \wac^2/9 +1}} \cr &
+ {\cos (3 \psi_0 + \arctan(4 \wac /3) -\arctan(2 \wac/3))
\over \sqrt{16 \wac^2/9 +1}} \cr &
+ {\cos  (3 \psi_0 - \arctan(4 \wac /3) -\arctan(2 \wac/3))
\over \sqrt{16 \wac^2/9 +1}}  \cr &
+ {\cos (3 \psi_0)
\over  \sqrt{4 \wac^2/9 +1}} \biggr). \cr}
\eqno(32)
\en

\noindent
After some algebraic manipulations this becomes

\eq\eqalign{
i_{1/3,1/3} = &\,  {J_{0}^{2}(\ad) \, J_1(\ad) \over 4 (4 \wac^2/9 +1)} \,
\biggl [\biggl ({3 - 8 \wac^2/9 \over (16 \wac^2/9
+1)}  + {2 \wac^2 /9 + 3/2 \over (4 \wac^2/9 +1)} \biggr) \cr &
\cos (3 \psi_0)
+  \biggl ({2 \wac/3 \over (4 \wac^2/9 +1)}
- {2 \wac /3 \over (16 \wac^2 /9 +1)} \biggr) \sin (3 \psi_0)
 \biggr]. \cr}
\eqno(33)
\en

In the limit of high frequencies this is

\eq
i_{1/3,1/3} = {81 \over 128 \, \wac^3} \, J_{0}^{2}(\ad) \, J_{1}(\ad)
\, \sin (3 \psi_0),
\eqno(34)
\en
and with our {\it ansatz} for $\psi_0$
\eq
\Delta i_{1/3,1/3} = {81 \over 128 \, \wac^3}
\, J_{0}^{2}(\ad) \, J_{1}(\ad)
\eqno(35)
\en
\noindent
whereas in the low-frequency limit we find

\eq
i_{1/3,1/3} = {9 \over 8} \,J_{0}^{2}(\ad) J_{1}(\ad) \, \cos(3 \psi_0).
\eqno(36)
\en
\bigskip
\noindent
{\bf (b2.2) $f=1/3, \nu = 2/3$}
\medskip
The number of modes to be included in the calculation keeps
growing, as promised: $(3,-2) = (1,0)_L + 2 \, (1,-1) + 0 \, (0,1)$,
$\, (3,-2) = (1,0)_L + 2 \, (1,0) -2 \, (0,1)$, $\, (3,-2) = (1,0)_L
+ (2,-1) - (0,1)$, $\, (3,-2) = (1,0)_L + (2,0) - 2 \, (0,1)$,
$\, (3,-2) = (1,0)_L + (2,-2) + 0 \, (0,1)$,  $\, (3,-2) = (1,0)_L +
(1,-1) + (1,0) - (0,1)$, and $\, (3,-2) = (1,0)_L + (1,-2) + (1,0)
+ 0 \, (0,1)$. In order to carry out the calculation we need
 $\, (1,-2) = (1,0)_L -2 \, (0,1)$, and $\, (2,-2) = (1,0)_L
+ (1,-1) - (0,1) = (1,0)_L + (1,0) -2 \, (0,1) = (1,0)_L + (1,-2) +
 0 \, (0,1)$ .
The calculation is analogous to the one
done in the previous section.
In the high-frequency limit we obtain:

\eq\eqalignno{
i_{1/3,2/3} =  {81 \over 128 \, \wac^3} \, &  J_{0}(\ad) \, \bigl (J_{1}^{2}
(\ad) \cr & + \,
J_0 (\ad) J_2 (\ad)/8 \bigr )
\cos(3 \psi_{0}),&(37) \cr}
\en
and after the usual assumption for $\psi_0$
\eq\eqalignno{
\Delta i_{1/3,2/3} =  {81 \over 128 \, \wac^3}  \, & J_0 (\ad) \, \bigl
(J_{1}^{2}(\ad) \cr & + \,
J_0 (\ad) J_2 (\ad)/8 \bigr ). &(38)\cr}
\en
\noindent
We have included higher-order corrections in Bessel functions
that this time yield a non-negligible contribution.
Once again, the step width decays as $1/\wac^3$.
 The low-frequency limit yields

\eq
i_{1/3,2/3} = {9 J_{0}(\ad) \over 8} \, \biggl(J_{1}^{2} (\ad) +
J_{0} (\ad) J_{2} (\ad) \biggr) \sin(3 \psi_{0}).
\eqno(39)
\en

\noindent
We get the expected single-junction behavior. \par

We see that at high frequencies the step-widths for
$f=1/3$  decrease
like $1/\wac^3$. The terms that vary like $1/\wac^2$  cancel out.
Compare this
result with the ones obtained for $f=1/2$. In those
cases we found that the step width varied like $1/\wac^2$. The
term varying like
$1/\wac$ also did cancel out. This result seems to be quite
general, and
there is a way of understanding it.
 The fractional steps in the present case correspond to the
subharmonic steps for a single junction. We can say that
the vertical supercurrent plays the role of an effective additional
degree of freedom, analogous to the role played by the capacitive
term for a single junction, in which case subharmonic steps
do appear. At high frequencies
the vertical supercurrent becomes negligible, and then the
equation for current conservation effectively becomes the equation
for an overdamped single junction, for which the subharmonic
steps are  absent.\par
Another observation that can be made at this point is that in
all the cases that  have been studied, the expressions of
 the step widths for $f=p/q$ and $\, \nu=n/q$ involve the
products of $q$ Bessel functions of $\, (\ad)$ such that
the sum or the difference of their orders is equal to $n$.
This is in accord with
Eq.~(16). The  pre-factor has a different dependence on the
frequency due to cancellations occuring among the different
modes.\par
\bigskip
\noindent
{\bf IV. NUMERICAL RESULTS}
\medskip
We use the same method of numerical integration used in reference $8$.
We briefly review it here, for convenience. The equations for current
conservation can be written as a matrix equation
\eq
{\bf M}_{ij} {d \theta_{j} \over dt} = F(\{\theta_{i'} - \theta {i} \})
\eqno(40)
\en
where $i'$ denotes a nearest neighbor to $i$. The matrix ${\bf M}$ is then
inverted yielding a set of coupled first-order differential equations
which we integrate using the fourth-order Runge-Kutta method.$^9$
The current
was uniformly injected at the left  boundary of an
$N \times N$ array. Furthermore, we used
periodic (with period q) boundary conditions in the direction
perpendicular to that  of injected the current. We normally used arrays
of size $N=3 q$ to $5 q$, in order to avoid effects due to the boundary
conditions. We used staircase configurations as initial
states in all of our simulations.
We restricted our observations to the cases $f=1/2$ and
$f=1/3$, for the already mentioned reasons.\par

Fig.~1 shows the power spectrum of the oscillating pieces corresponding
to the gauge-invariant phase differences on horizontal and vertical bonds
for $f=1/2$, away from any step. The slope of the linear term on
horizontal bonds agrees with the value of the Josephson frequency
observed
in the power spectrum.  Peaks 3 though 6 are absent on vertical bonds,
in particular notice the absence of the $(0,1)$ mode.
This is accord with
our assumptions. This same behavior has been observed for
$f=1/3$. Peak 4
in  Fig.~1 (a) corresponds to the $(2,-1)$ mode.
On the first fractional step the frequency corresponding to this
mode is zero, and the corresponding supercurrent yields the
additional DC current on the step; the same is true for Peak 7
on the first integer step. Also, on the latter step, the
mode corresponding to Peak 1 corresponds to the zero frequency
mode.
On the different steps we  get the same picture for
the power spectrum, with the difference that in these cases the
motion is periodic.\par

Fig.~2 displays the behavior of the gauge-invariant phase differences
on horizontal and vertical bonds for $f=1/2$ and $\nu=1$. The presence
of a zero-frequency (spatially varying DC component of the phase)
is clear. This is also in agreement with our results.\par

In Fig.~3 we show the different step widths as a function of
 $\iac$ for  $\wac=2$, and $\wac=3$. At higher frequencies
the fractional steps become suppressed. These results are in agreement
with the simulational results obtained by other groups.$^{3,5}$\par

In Fig.~4 we show the variation of the step widths with $\iac$
 for $\, \wac = 2.0$, and $\, \wac=3.0$.
The qualitative behavior of the step widths is the
same that was found for $f=1/2$.
\bigskip
\noindent
{\bf ACKNOWLEDGEMENTS}
\medskip
The author wishes to thank his thesis advisor Thomas Halsey
for helpful discussions and for his encouragement. S. J.
Lee also provided some helpful comments. This work
was completed in partial fulfillment of the Ph. D requirements
at the University of Chicago. This research was supported by the
Materials Research Laboratory at the University of Chicago.

%\bigskip
%\noindent
%{\bf APPENDIX A}
%\medskip
%\vfill
%\eject
\bigskip
\noindent
{\bf REFERENCES}
\medskip
\item{1.} S. Shapiro, Phys. Rev. Lett. {\bf 11}, 80 (1963);
B. D. Josephson, Phys. Lett. {\bf 1}, 251 (1962).

\item{2.} S. P. Benz, M. S. Rzchowski, M. Tinkham, and C. J. Lobb,
Phys. Rev. Lett. {\bf 64}, 693 (1990).

\item{3.} K. H. Lee, D. Stroud, and J. S. Chung, Phys. Rev. Lett.
 {\bf 64}, 962 (1991); K. H. Lee and D. Stroud, Phys. Rev. B {\bf 43},
5280 (1991).

\item{4.} M. S. Rzchowski, L. L. Sohn, and M. Tinkham, Phys. Rev. B
{\bf 43}, 8682 (1991).

\item{5.} M. Octavio, J. U. Free, S. P. Benz, R. S. Newrock, D. B. Masta
and C. J. Lob, Phys. Rev. B {\bf 44}, 4601 (1991); U. Free, S. P. Benz,
M. S. Rzchowski, M. Tinkham, C. J. Lobb, and M. Octavio, Phys. Rev.
B {\bf 41}, 7267 (1990).

\item{6.} S. J. Lee and T. C. Halsey, Phys. Rev. B {\bf 47}, 5133 (1993).

\item{7.} T. C. Halsey, Phys. Rev. B {\bf 41}, 11634 (1990).

\item{8.} I. Marino and T. C. Halsey, Phys. Rev. B {\bf 50}, 6289 (1994).

\item{9.} For a discussion of the Runge-Kutta method of integration,
see, for instance, W. H. Press, B. P. Flannery, S. A. Teukolky, and
W. T. Vetterling, {\it Numerical Recipes: The Art of Scientific Computing}
(Cambridge Univ. Press, Cambridge, 1986), Chapter 15.

\vfill
\eject
\bigskip
\noindent
{\bf FIGURE CAPTIONS}
\item{1.} (a) Power spectrum $\, P(\omega)$ of horizontal
 phase oscillations for $f=1/2$
away from any step. The  peaks  $(1-6)$
correspond to  the frequencies $\, (\wac - \wj)$, $\, \wj$,
 $\, \wac$,
 $\, (2 \wj - \wac)$, $\, 2 \wj$, and $\, (2 \wac - 2 \wj)$ respectively;
(b) Power spectrum of vertical phase oscillations. Notice the
absence of  peaks 3-6, which is in accord with the selection rule
for the existence of modes on vertical bonds
 (see the remark following Eq.~(10)).
\item{2.} Gauge-invariant phase differences on two
vertical bonds lying on the same column for $f=1/2$
on the first integer step. They are out of phase by $\, \delta = \pi$,
as assumed in our work. The average value
of the phase oscillations is non-vanishing and is depicted
by a solid line in the figures. This represents the zero-frequency
mode, which changes phase by $\, \pi$ when translated by
one lattice spacing on the array.

\item{3.} Step widths for $f=1/2$ and $\, \nu=1/2, \, 1,\, 3/2$, and $2$
for
(a)  $\, \wac = 2.0$; (b) and $\, \wac= 3.0$.
Fractional steps are suppressed relative to the integer steps.

\item{4.} (a) Step widths for $f=1/3$, $\, \nu = 1/3, \, 2/3, \, 1$ , and
$\, \wac=2.0$; (b) The frequency is $\, \wac=3.0$. The
steps are considerably smaller than those for $f=1/2$.

\vfill
\eject
\end